\documentclass[12pt,dvips]{article}
\usepackage{amsmath,amssymb}
\usepackage{pstricks,pst-node}
\makeatletter

\@addtoreset{equation}{section}
\makeatother

\def\cob{\delta}

\def\hf{\frac{1}{2}}

\def\til#1{\widetilde{#1}}

\def\si{\sigma}

\def\lap{\Delta}

\def\ri{\right}
\def\riya{\rightarrow}

\def\la{\lambda}
\def\La{\Lambda}

\def\al{\alpha}

\def\rt#1{\sqrt{#1}}

\def\sitarel#1#2{\mathrel{\mathop{\kern0pt #1}\limits_{#2}}}

\newcommand{\nn}{\nonumber \\}
\newcommand{\tsi}{\widetilde{\sigma}}

\begin{document}
\baselineskip 0.8cm
\vspace*{2.5cm}

\begin{center}
{\LARGE\bf A Note on the Partition Function of}\\
\vskip1mm
{\LARGE\bf ABJM theory on $S^3$}
\end{center}
\vskip 8ex
\begin{center}
  Kazumi Okuyama\\
  {\it Department of Physics, Shinshu University, Matsumoto 390-8621, Japan} 
  {\tt kazumi@azusa.shinshu-u.ac.jp}
\end{center}
%%%%%%%%%%%%%%%%%%%%%%%
\vskip 25mm
%%%%%%%%%%%%%%%%%%%%%%%
%
\baselineskip=6mm
We study the partition function $Z$ of $U(N)_k\times U(N)_{-k}$
Chern-Simons matter theory (ABJM theory) on $S^3$ which is
recently obtained by the localization method.
We evaluate the eigenvalue integral in $Z$
exactly for the $N=2$ case. 
We find that $Z$ has a different dependence
on $k$ for even $k$ and odd $k$.
%If $k$ is odd,  $Z$ is written as a sum of two parts, which we call
%the bulk part and the orbifold part.
%If $k$ is even, $Z$ has only the orbifold part. 
We comment on the possible implication of this result in
the context of AdS/CFT correspondence.
\vskip 50mm
\noindent

\newpage
%\tableofcontents
\section{Introduction}
In the seminal paper \cite{Aharony:2008ug},
the theory on the $N$ coincident  M2-branes
on the orbifold $\mathbb{R}^8/\mathbb{Z}_k$
was identified as
the $d=3$ ${\cal N}=6$ $U(N)_k\times U(N)_{-k}$
Chern-Simons matter theory (ABJM theory).
Recently, 
the partition function 
$Z_{N,k}$ of ABJM theory
on $S^3$ was obtained by the localization method \cite{Kapustin:2009kz},
and $Z_{N,k}$ was given
in the form of a matrix integral.
The behavior of $Z_{N,k}$ 
has been analyzed previously 
\cite{Drukker:2010nc,Drukker:2011zy,Fuji:2011km,Santamaria:2010dm} 
in the 't Hooft limit
\begin{equation}
k,N\riya \infty, \quad t=\frac{N}{k}={\rm fixed}~,
\label{tHooft}
\end{equation}
and it was shown that 
the free energy $F=-\log Z_{N,k}$ 
exhibits the correct $N^{\frac{3}{2}}$ scaling
as predicted by the holographic dual gravity theory.
The ABJM theory in the 't Hooft limit
is holographically dual to 
the type IIA theory on ${\rm AdS}_4\times\mathbb{CP}^3$,
which appears from the $S^1$ reduction
of the M-theory on ${\rm AdS}_4\times S^7/\mathbb{Z}_k$
when $k\gg N^{\frac{1}{5}}$.

However, if we are interested in
the  dynamics of M2-branes in the truly M-theory regime, or
in the strong coupling regime of
type IIA theory,  we need to know the behavior of
$Z_{N,k}$ at finite $k$, since the IIA string coupling
is inversely proportional to $k$.
Of particular interest is the ABJM theory at $k=1$, which is conjectured
to describe the M2-branes on the flat eleven dimensional Minkowski space.
Therefore we might want to develop a technique to analyze
the partition function $Z_{N,k}$
in the M-theory regime where
\begin{equation}
N\riya\infty,\quad k={\rm finite}~.
\end{equation}
This regime was studied 
in \cite{Herzog:2010hf} by the saddle point method for the eigenvalue
integral.

In this paper we find the exact partition function $Z_{N,k}$ for $N=2$
with finite $k$ by
performing the eigenvalue integral explicitly for the $N=2$ case.
We find that the result depends on the parity of $k$:
\begin{eqnarray}
Z_{2,\,{\rm odd}\,k}
&=&
\frac{1}{k}\sum_{s=1}^{k-1}(-1)^{s-1}\left(\hf-\frac{s}{ k}\right)
\tan^{2}\frac{\pi s}{ k}+\frac{(-1)^{\frac{k-1}{2}}}{\pi}~,\\
Z_{2,\,{\rm even}\,k}&=&\frac{1}{ k}\sum_{s=1}^{k-1}(-1)^{s-1}\left
(\hf-\frac{s}{ k}\right)^2\tan^2\frac{\pi s}{ k}~.
\end{eqnarray}
For both even $k$ and odd $k$ cases,
the summation over $s$ has a natural interpretation as the
effect of  $\mathbb{Z}_k$ orbifolding of
$\mathbb{R}^8/\mathbb{Z}_k$.
%When $k$ is odd, $Z_{2,k}$ has the second term 
%$\frac{(-1)^{\frac{k-1}{2}}}{\pi}$,
%which we call the bulk part since this term is not suppressed by $\frac{1}{ k}$.

This paper is organized as follows.
In section 2, we first rewrite the
partition function of ABJM theory on $S^3$
in terms of the integrals associated with
the cyclic permutations. 
Then we consider the grand partition function
of ABJM theory,
following the similar analysis of the matrix integrals 
which arise from the
dimensional reduction of 
super Yang-Mills theories to 0-dimension \cite{Moore:1998et,Kazakov:1998ji}.
We also comment on the mirror description
of the partition function of ABJM theory.
In section 3, we compute the partition function of
$U(2)_k\times U(2)_{-k}$ ABJM theory and find that the
result depends on the parity of $k$. 
In section 4,
we speculate the possible implication of this result
in the context of AdS/CFT correspondence.
In Appendix A and B, we present the details of the calculation
of integrals used in section 3.

\section{Structure of the Partition Function of ABJM Theory on $S^3$}
\subsection{Grand Partition Function of ABJM Theory}
Recently, by applying the localization method 
of \cite{Pestun:2007rz},
the partition function of 
general ${\cal N}=2$ Chern-Simons matter
theories on $S^3$ with the gauge group $G$
and the matter chiral multiplet
in a representation $R\oplus R^*$
was obtained in a form of matrix integral \cite{Kapustin:2009kz}
\footnote{
The partition function of the theory
with matter multiplet in the
non-self-conjugate representation
was obtained in 
\cite{Jafferis:2010un,Hama:2010av}. 
Note that (2.1) is valid only when
the R-charge carried by the matter multiplet is $1/2$. 
The partition function for the case of
non-canonical R-charge $q$ was also calculated 
in \cite{Jafferis:2010un,Hama:2010av}. 
See also \cite{Marino:2011nm} for a recent
review on the localization technique in $d=3$ theories.}
\begin{equation}
Z=\frac{1}{|W|}\int da \,e^{-i\pi k a^2}\frac{\det_{Ad}(\sinh\pi a)
}{\det_R(\cosh\pi a)}~.
\end{equation}
Above, the integral of $a$ is over the Cartan subalgebra
of $G$, $|W|$ is the order of the Weyl group of $G$, and
$k$ is the Chern-Simons coupling which is quantized to be an integer.  
Note that $a$ originates from the constant mode of the real scalar field
in the vector multiplet.

Since the ABJM theory is the $d=3$ $U(N)_k\times U(N)_{-k}$
Chern-Simons theory
with bi-fundamental matter multiplets, its partition function on $S^3$ is
given by
%\begin{equation}
%Z_{N,k}={1\o(N!)^2}\int {d^N\si d^N\tsi\o(2\pi)^{2N}} 
%\,e^{{ik\o4\pi}(\si^2-\tsi^2)}\,{\prod_{i<j}\sinh^2{\si_i-\si_j\o2}
%\sinh^2{\tsi_i-\tsi_j\o2}\o\prod_{i,j}
%\cosh^2{\si_i-\tsi_j\o2}}
%\end{equation} 
%By rescaling the integration variables 
%$\si_i,\tsi_i\riya 2\pi \si_i,2\pi \tsi_i$, we get
\begin{equation}
Z_{N,k}=\frac{1}{(N!)^2}\int d^N\si d^N\til{\si}
\,\lap(\si,\til{\si})^2\,e^{i\pi k(\si^2-\til{\si}^2)} ~,
\label{Zmat}
\end{equation}
where $\si^2$ is the shorthand for $\sum_{i=1}^N\si_i^2$, 
and similarly $\tsi^2=\sum_{i=1}^N\tsi_i^2$, and
$\lap(\si,\tsi)$ is given by
\begin{equation}
\lap(\si,\tsi)=
\frac{\prod_{i<j}\sinh\pi(\si_i-\si_j)\sinh\pi(\tsi_i-\tsi_j)}{\prod_{i,j}
\cosh\pi(\si_i-\tsi_j)}~.
\end{equation}
Using the Cauchy identity \cite{Kapustin:2010xq}
\begin{equation}
\lap(\si,\tsi)=\sum_{\rho\in S_N}(-1)^\rho
\prod_{i=1}^N\frac{1}{ \cosh\pi(\si_i-\tsi_{\rho(i)})}~,
\end{equation}
the partition function is rewritten as
\begin{equation}
Z_{N,k}=
\frac{1}{ N!}\int{d^N\si d^N\tsi} \,e^{i\pi k(\si^2-\tsi^2)}
\sum_{\rho\in S_N}(-1)^\rho\prod
_{i=1}^N\frac{1}{ \cosh\pi(\si_i-\tsi_i)\cosh\pi(\si_i-\tsi_{\rho(i)})}~.
\label{Zint}
\end{equation}

The sum over permutations 
can be simplified by noting that 
the integral depends only on the conjugacy class of
permutation.
The conjugacy class of permutation $\rho$ 
is labeled by the cycle of length $\ell$ and the number 
$d_\ell$ of such cycles
contained in $\rho$
\begin{equation}
[\rho]=[1^{d_1}2^{d_2}\cdots N^{d_N}]\equiv
\Big[\prod_\ell \ell^{d_\ell}\Big],\quad
N=\sum_\ell \ell d_\ell~.
\end{equation}
The number of elements in the conjugacy class
$[\rho]$ and the signature
are given by
\begin{equation}
\# [\rho]=\frac{N!}{ \prod_\ell \ell^{d_\ell}d_\ell!},\quad
(-1)^\rho=(-1)^{\sum_\ell d_\ell(\ell-1)}~.
\end{equation}

One can show that the integral
in (\ref{Zint}) is decomposed into the integral
associated with the cyclic permutation
\begin{equation}
Z_{N,k}=\sum_{d_\ell\geq0,\sum\ell d_\ell=N}
\prod_{\ell=1}^N \frac{1}{ d_\ell!}\left[\frac{(-1)^{\ell-1}A_{\ell,k}}{\ell}\right]^{d_\ell}~,
\label{Zincycle}
\end{equation}
where $A_{\ell,k}$ denotes the integral coming from the
cycle of length $\ell$
\begin{equation}
A_{\ell,k}=
\int{d^\ell\si d^\ell\tsi} \,e^{i\pi k(\si^2-\tsi^2)}
\prod_{i=1}^\ell\frac{1}{ \cosh\pi(\si_i-\tsi_i)\cosh\pi(\si_i-\tsi_{i+1})}~.
\label{A-ell}
\end{equation}
Here the mod-$\ell$
identification $\tsi_{\ell+1}\equiv\tsi_1$ should be understood.

By introducing the chemical potential $\mu$
for $N$, the grand partition function is defined by
\begin{equation}
{\cal Z}_k(\mu)=\sum_{N=0}^\infty e^{\mu N}Z_{N,k}~.
\end{equation}
From (\ref{Zincycle}) one can easily see that 
${\cal Z}_k(\mu)$ is exponentiated after summing over $d_\ell$'s
\begin{equation}
{\cal Z}_k(\mu)=\exp\left[
\sum_{\ell=1}^\infty \frac{(-1)^{\ell-1}}{ \ell}e^{\mu\ell} A_{\ell,k}\right]~.
\label{grandZ}
\end{equation}
Once we know the grand partition function, we can recover the
fixed $N$ partition function from the integral
of ${\cal Z}_k(\mu)$
by analytically continuing the chemical potential 
to a pure imaginary value $\mu=i\theta$
\begin{equation}
Z_{N,k}=\int_0^{2\pi}\frac{d\theta}{2\pi }e^{-iN\theta}{\cal Z}_k(i\theta)~.
\end{equation}
It would be interesting to see whether the grand partition function
of ABJM theory has a hidden integrable structure as in \cite{Kazakov:1998ji}.

\subsection{Mirror Description of ABJM Theory}
By the mirror symmetry, the ABJM theory is
dual to a theory without Chern-Simons term.
More concretely, the mirror of ABJM theory is a $U(N)$ super Yang-Mills theory with
matter hypermultiplets in certain representations of $U(N)$.
As discussed in \cite{Kapustin:2010xq},
the partition function on $S^3$ is a useful tool to check this
type of mirror symmetry.
The key relation to prove the equality of partition functions
of the original theory and its mirror is the following identity
\begin{equation}
\int dx \frac{e^{2\pi i x\si}}{\cosh\pi x}=\frac{1}{\cosh\pi\si}~.
\label{cosh-id}
\end{equation}
Using this relation, the partition function of ABJM theory
$Z_{N,k}$ is rewritten as
\begin{equation}
Z_{N,k}=\frac{k^{2N}}{ N!}
\int d^N\si d^N\tsi d^Nx d^Ny\,
\sum_{\rho\in S_N}(-1)^\rho
\frac{e^{i\pi k\sum_{i=1}^N[\si_i^2-\tsi_i^2+2x_i(\si_i-\tsi_i)
+2y_i(\si_i-\tsi_{\rho(i)})]}}
{\prod_{i=1}^N\cosh\pi kx_i\cosh\pi ky_i}~.
\label{int-mirrorZ}
\end{equation}
After doing the Gaussian integral for $\si,\tsi$
and using the identity (\ref{cosh-id}) again for the $y$-integral,
(\ref{int-mirrorZ}) becomes
\begin{equation}
Z_{N,k}=\frac{1}{ N!}\int \prod_{i=1}^N dx_i
\sum_{\rho\in S_N}(-1)^\rho\prod_{i=1}^N\frac{1}{\cosh \pi k x_i\cosh\pi (x_i-x_{\rho(i)})}~.
\end{equation}
Applying the Cauchy identity for the sum over permutations, we arrive 
at the mirror expression of the partition function of ABJM theory
\begin{equation}
Z_{N,k}=\frac{1}{ N!}\int \prod_{i=1}^N dx_i\,\frac{\prod_{i<j}\sinh^2\pi(x_i-x_j)}
{\prod_i\cosh\pi kx_i
\prod_{i,j}\cosh\pi(x_i-x_j)}~.
\label{mirrorZ}
\end{equation}
From this, we can read off the matter content of the mirror of
ABJM theory.
When $k=1$, the mirror theory is the $U(N)$ super Yang-Mills
theory with one adjoint and one fundamental hypermultiplets, 
where
the factors $1/\prod_{i,j}\cosh\pi(x_i-x_j)$
and $1/\prod_i\cosh\pi x_i$ in (\ref{mirrorZ}) are the 
1-loop determinant of those hypermultiplets, respectively
\cite{Kapustin:2010xq}. When $k\geq2$ 
it is not clear whether the factor 
$1/\prod_i\cosh k\pi x_i$ can be interpreted as the 
1-loop determinant of hypermultiplet in some representation $R$.
In particular 
it is different from the 1-loop
determinant of 
hypermultiplet in the $k^{\rm th}$ symmetric product of
fundamental representations.

The grand partition function of the mirror theory
of ABJM theory has the same form as
(\ref{grandZ}), and the contribution from the cycle
of length $\ell$ in the mirror description is given by
\begin{equation}
A_{\ell,k}=\int d^\ell x\prod_{i=1}^\ell \frac{1}{ \cosh\pi k x_i
\cosh\pi (x_i-x_{i+1})}~.
\end{equation}

\section{Partition function of $U(2)_k\times U(2)_{-k}$ ABJM theory}
In this section, we study the partition function $Z_{2,k}$ of
$U(2)_k\times U(2)_{-k}$ ABJM theory.
Since this model is conjecture to describe the dynamics of two M2-branes on 
$\mathbb{R}^8/\mathbb{Z}_k$,
we expect that some information of the two-body interaction of M2-branes
is contained in the partition function $Z_{2,k}$.
Therefore, 
the study of the partition function 
of $U(2)_k\times U(2)_{-k}$ theory
would be a modest first step toward the understanding of the still mysterious
multiple M2-brane dynamics.\footnote{In a slightly different context, 
the exact evaluation of the partition function of $U(2)$ IIB matrix model 
was reported in \cite{Suyama:1997ig,Green:1996xf}.}

Here we evaluate the eigenvalue integral 
of $Z_{N,k}$ in (\ref{Zint})  explicitly for the $N=2$ case. 
%\begin{align}
%Z_{2,k}={1\o4}\int & d\si_1d\si_2d\tsi_1d\tsi_2
%e^{i\pi k(\si_1^2+\si_2^2-\tsi_1^2-\tsi_2^2)}\nn
%&\times 
%{\sinh^2\pi(\si_1-\si_2)\sinh^2\pi(\tsi_1-\tsi_2)\o\cosh^2\pi(\si_1-\tsi_1)
%\cosh^2\pi(\si_2-\tsi_2)
%\cosh^2\pi(\si_1-\tsi_2)
%\cosh^2\pi(\si_2-\tsi_1)}~.
%\end{align}
To do that, we first
rewrite $Z_{2,k}$ as a combination of 
the integral $A_{\ell,k}$ coming from the cyclic permutation of length $\ell$ 
as shown in (\ref{Zincycle})
\begin{equation}
Z_{2,k}=\hf\Big[(A_{1,k})^2-A_{2,k}\Big]~.
\label{Ztwok}
\end{equation}
Although $A_{2,k}$ is originally written as an integral over four
variables (\ref{A-ell}), 
after some computation 
this four-variable integral can be
reduced to a single variable integral.
We find that $A_{1,k}$ and $A_{2,k}$ are given by (see Appendix A for details)
\begin{eqnarray}
A_{1,k}&=&\frac{1}{ k}~,\\
A_{2,k}&=&\int_{-\infty}^\infty d\la\frac{2\la}{\sinh\pi k\la\cosh^2\pi\la}
=\frac{1}{ k^2}-\int_{-\infty}^\infty d\la
\frac{2\la}{\sinh\pi k\la}\frac{\sinh^2\pi\la}{\cosh^2\pi\la}~.
\label{A-two}
\end{eqnarray}
Plugging this into (\ref{Ztwok}), we obtain
\begin{equation}
Z_{2,k}=\int_{-\infty}^\infty d\la \frac{\la}{\sinh\pi k\la}
\frac{\sinh^2\pi\la}{\cosh^2\pi\la}~.
\label{Ztwoint}
\end{equation}
Note that $\la$ is related to the original variables (up to permutation) as
\begin{equation}
\la=\si_1-\tsi_1~.
\end{equation}
As explained in Appendix B,
the remaining $\la$-integral can be evaluated by picking up the residues of the poles 
of $\frac{1}{\sinh\pi k\la}$ and $\frac{1}{\cosh^2\pi \la}$. 
It turns out that the result depends on the parity of $k$
\begin{eqnarray}
Z_{2,\,{\rm odd}\,k}
&=&
\frac{1}{ k}\sum_{s=1}^{k-1}(-1)^{s-1}\left(\hf-\frac{s}{ k}\right)
\tan^{2}\frac{\pi s}{ k}+\frac{(-1)^{\frac{k-1}{2}}}{\pi}~,\\
Z_{2,\,{\rm even}\,k}&=&\frac{1}{ k}\sum_{s=1}^{k-1}(-1)^{s-1}\left
(\hf-\frac{s}{ k}\right)^2\tan^2\frac{\pi s}{ k}~.
\label{result}
\end{eqnarray}
In the above expression of $Z_{2,\,{\rm even}\,k}$,
the $s=\frac{k}{2}$ term should be understood as the limit
\begin{equation}
\lim_{s\riya\frac{k}{2}}\frac{1}{ k}(-1)^{s-1}\left(\hf-\frac{s}{ k}\right)^2
\tan^2\frac{\pi s}{ k}=\frac{(-1)^{\frac{k}{2}-1}}{ k\pi^2}~.
\end{equation} 

Let us consider the physical interpretation of this result (\ref{result}).
For both even $k$ and odd $k$ cases, the sum over $s$
comes from the poles at $\sinh\pi k\la=0$.
It is natural to interpret
this sum as the effect of the $\mathbb{Z}_k$ orbifolding of 
$\mathbb{R}^8/\mathbb{Z}_k$.
On the other hand, the second term $\frac{(-1)^{\frac{k-1}{2}}}{\pi}$ in 
$Z_{2,\,{\rm odd}\,k}$ comes from the pole at $\cosh\pi\la=0$.
This pole corresponds to the zero of the 1-loop determinant of
the bi-fundamental hypermultiplet, so it represents
a singularity on the space of vector multiplet
scalar fields where one of the bi-fundamental 
hypermultiplet becomes massless.
However, the location of the singularity is 
at the imaginary value of the scalar field
\begin{equation}
\si_1-\tsi_1=\frac{i}{2}~,
\end{equation}
and hence this singularity is not realized in the physical theory.
We should also mention that the poles coming from the 
$\frac{1}{\sinh\pi k\la}$ 
factor
do not correspond to the zeros of the 1-loop determinant
of the
hypermultiplets in the original ABJM theory.
Those poles effectively show up only after integrating out some
of the variables $\si_i,\tsi_i$, which are coupled via
the Chern-Simons term $e^{\pi ik(\si^2-\tsi^2)}$.

From (\ref{A-two}), we see that $A_{2,k}$ is positive.
Therefore, we find the inequality
\footnote{
The normalization of the partition function in \cite{Drukker:2010nc}
is
different from ours by the factor of 2
in the 1-loop determinant. Namely, 
the partition function in \cite{Drukker:2010nc}
is related to ours by the replacement
$\sinh\riya 2\sinh, \cosh\riya2\cosh$
\begin{equation}
Z^{({\rm DMP})}_{N,k}=\frac{1}{(N!)^2}\int {d^N\si d^N\tsi} 
\,e^{i\pi k(\si^2-\tsi^2)}
\,\left[\frac{\prod_{i<j}2\sinh\pi(\si_i-\si_j)
\cdot2\sinh\pi(\tsi_i-\tsi_j)}{\prod_{i,j}
2\cosh\pi(\si_i-\tsi_j)}\ri]^2~.
\end{equation}
One can easily see that the 
difference between $Z^{({\rm DMP})}_{N,k}$ and ours is
just the overall factor $2^{-2N}$
\begin{equation}
Z^{({\rm DMP})}_{N,k}=2^{-2N}Z^{({\rm ours})}_{N,k}~.
\end{equation}
However, this factor drops out when taking the ratio of
$(Z_{1,k})^2$  and $Z_{2,k}$
\begin{equation}
\frac{(Z^{({\rm DMP})}_{1,k})^2}{Z^{({\rm DMP})}_{2,k}}
=\frac{(Z^{({\rm ours})}_{1,k})^2}{Z^{({\rm ours})}_{2,k}}~.
\end{equation}
Therefore, the statement $Z_{2,k}<\hf (Z_{1,k})^2$ has a physical meaning
regardless of the normalization we choose.
}
\begin{equation}
Z_{2,k}<\hf(Z_{1,k})^2
\label{ineqZ}
\end{equation}
where $Z_{1,k}=A_{1,k}=\frac{1}{ k}$ is the partition function
of $U(1)_k\times U(1)_{-k}$ theory. 
From  this inequality (\ref{ineqZ}), it is tempting to
draw a conclusion  that the
binding energy of two M2-branes is negative and 
M2-branes tend to dissociate into a configuration of two separated
M2-branes. However, we think this is not the correct interpretation.
When the ABJM theory is put on $S^3$, the bi-fundamental matter
multiplets acquire a mass term from the coupling to the
curvature of $S^3$,
and hence the moduli space corresponding to the freely
moving M2-branes on $\mathbb{R}^8/\mathbb{Z}_k$ is lifted.
Therefore, the free energy of ABJM theory on $S^3$ is not a suitable
measure of the binding energy of M2-branes on flat 
$\mathbb{R}^{1,2}\times \mathbb{R}^{8}/\mathbb{Z}_k$.
Rather, the partition function on $S^3$ is a natural
quantity to consider
in the context of the Euclidean version of AdS/CFT duality, where $S^3$ 
appears as the boundary of Euclidean ${\rm AdS}_4$.
In the next concluding section we discuss a possible implication of our result
in the context of ${\rm AdS}_4/{\rm CFT}_3$ duality.

\section{Discussions}
As discussed in \cite{Aharony:2008ug},
the ABJM theory is dual to the M-theory on 
${\rm AdS}_4\times S^7/\mathbb{Z}_k$ with the metric
\begin{equation}
ds^2=\frac{R^2}{4}ds^2_{{\rm AdS}_4}+R^2ds^2_{S^7/\mathbb{Z}_k}~,
\end{equation}
where the radius of curvature $R$ is given by
\begin{equation}
\left(\frac{R}{ l_{p}}\right)^6=32\pi^2kN~.
\end{equation}
The classical $d=11$ supergravity description is valid when
the radius of $S^7/\mathbb{Z}_k$ is much larger than the 
eleven-dimensional Planck length
$l_p$
\begin{equation}
l_p\ll \frac{R}{ k}~~\riya~~k^5\ll N~.
\label{Mregime}
\end{equation}
In particular,
the large $N$ limit of ABJM theory with $k$
fixed to a finite integer is in the regime of
(\ref{Mregime}).

On the ABJM theory side, 
it seems that the even/odd $k$ difference 
of the behavior of the partition function $Z_{N,k}$ persists for $N>2$.
This is because,
in the integral of $A_{\ell,k}$ in (A.6),
the pole of the form $\frac{1}{\sinh\pi k\la}$
related to the $\mathbb{Z}_k$ orbifolding
appears also for general  $\ell>2$ in the same way as $A_{2,k}$
by integrating out some of the variables in $\si_i,\tsi_i$
coupled through the Chern-Simons term,
and the remaining integral over $\la$ depends on the parity of $k$.
Since the partition function $Z_{N,k}$ is written as a combination
of $A_{\ell,k}$ (\ref{Zincycle}), 
$Z_{N,k}$ also depends on the parity of $k$,
unless some miraculous cancellation happens. 
But we think that is unlikely and the dependence on the 
parity of $k$ is not an artifact of $Z_{2,k}$ but the general property of
$Z_{N,k}$ for all $N\geq2$.

If we believe in the duality between the
ABJM theory and M-theory on ${\rm AdS}_4\times S^7/\mathbb{Z}_k$,
this difference of even/odd $k$
must be encoded in the M-theory dual, perhaps in a very 
subtle way. However, so far there is no known indication of this difference
in the supergravity approximation of  
M-theory on ${\rm AdS}_4\times S^7/\mathbb{Z}_k$.
Even if we take into account of the wrapped brane configuration
in this background, 
the bulk theory seems to be insensitive to the parity of $k$.
In fact, the BPS configuration of
M5-branes wrapped on the 3-cycle in $S^7/\mathbb{Z}_k$
is characterized by the homology class
\begin{equation}
H_3(S^7/\mathbb{Z}_k)=\mathbb{Z}_k~,
\end{equation}
which is interpreted as the fractional 
M2-brane charge \cite{Aharony:2008gk}.
Clearly, this charge does not distinguish the parity of $k$.
It might be the case that the even/odd $k$ difference 
appears in the bulk theory as 
some sort of quantum effects in M-theory, which cannot 
be seen in the supergravity approximation.
If this is true,
it would be nice to understand this effect better.

In the regime where
\begin{equation}
N^{\frac{1}{5}}\ll k\ll N~,
\end{equation}
the bulk theory is described by
the type IIA 
string theory on 
${\rm AdS}_4\times\mathbb{C}\mathbb{P}^3$.
%\begin{equation}
%g_s={2\pi i\o k}~.
%\end{equation}
On the CFT side, this regime is related to the 't Hooft limit
of ABJM theory (\ref{tHooft}),
%\begin{equation}
%k,N\riya \infty, \quad t={N\o k}={\rm fixed}~,
%\label{tHooft}
%\end{equation}
and the classical type IIA supergravity description becomes good
when 
the 't Hooft coupling $t=\frac{N}{ k}$ is large.
When comparing the
free energy $F=-\log Z_{N,k}$ of ABJM theory and the 
classical action of the bulk
supergravity theory, we need to perform an 
analytic continuation of $Z_{N,k}$ as a function of $k$ and $N$.
In particular, when determining the eigenvalue distribution
for the matrix integral (\ref{Zmat})
in the 't Hooft limit, the analytic continuation in
$k$ is implicitly assumed.

Our result suggests that the analyticity in $k$
is not obvious a priori, even in the large $N$ regime.
In some cases of Chern-Simons-matter theories, the analytic continuation in $k$ requires the deformation
of integration contour.
However, the integral representation of $Z_{2,k}$ in 
(\ref{Ztwoint}) is well-defined for $k\in\mathbb{R}$ without
changing the integration contour of $\la$.
From this integral representation (\ref{Ztwoint}),
one can see that $Z_{2,k}$ 
decreases monotonically as a function of 
$k$ \footnote{We would like to thank
the referee of Prog. Theor. Phys. for pointing this out.},
and the expression (\ref{Ztwoint}) for $k\in\mathbb{R}$ serves as an interpolating function
of our result (\ref{result}) for integer $k$.
It would be nice to see if
similar analytic continuation is possible for $N>2$ 
without deforming the integration contour.

\vskip10mm
\centerline{\bf Acknowledgements}
\noindent
I would like to thank Kazuo Hosomichi for a
nice lecture on the localization 
method in supersymmetric gauge theories at Chubu Summer School 2011.
This work is supported in part by JSPS Grant-in-Aid for Young Scientists
(B) 23740178.

\vskip10mm
%%%%%%%%%%%%% appendix %%%%%%%%%
\appendix
%\noindent
%{\bf\LARGE Appendix}
\section{Computation of $A_{\ell,k}$}
In this Appendix, by performing the integration of two variables,
we rewrite 
the $2\ell$-variable integral $A_{\ell,k}$ given in 
(\ref{A-ell}) into 
the integral of $2(\ell-1)$ variables.
Using this expression, we find $A_{1,k}=\frac{1}{k}$. We 
also find the
expression of $A_{2,k}$ as a single variable integral. 
\subsection{Writing $A_{\ell,k}$ as the integral of $2(\ell-1)$ variables}
For readers convenience, we repeat the integral $A_{\ell,k}$
in (\ref{A-ell})
\begin{equation}
A_{\ell,k}=\int d^\ell\si d^\ell\tsi\,e^{i\pi k(\si^2-\tsi^2)}\prod_{i=1}^\ell
\frac{1}{\cosh\pi(\si_i-\tsi_i)\cosh\pi(\si_i-\tsi_{i+1})}~.
\end{equation}
This integral can be simplified by the 
following change of variables
\begin{equation}
(\si_1,\cdots,\si_\ell,\tsi_1,\cdots,\tsi_{\ell})
\riya(\la_1,\cdots,\la_\ell,\til{\la}_1,\cdots,\til{\la}_{\ell-1},\tsi_\ell)
\end{equation}
where
\begin{equation}
\la_i=\si_i-\tsi_i~~(i=1,\cdots,\ell)~,\qquad
\til{\la}_i=\si_i-\til{\si}_{i+1}~~(i=1,\cdots,\ell-1)~.
\label{ladef}
\end{equation}
In terms of these new variables, the integral becomes
\begin{eqnarray}
A_{\ell,k}&=&\int d^\ell\la d^{\ell-1}\til{\la}d\tsi_\ell
\prod_{i=1}^{\ell}\frac{1}{ \cosh\pi\la_i}
\prod_{i=1}^{\ell-1}\frac{1}{\cosh\pi\til{\la}_i}
\cdot\frac{1}{\cosh\pi(\sum_{i=1}^\ell\la_i-\sum_{i=1}^{\ell-1}\til{\la}_i)}\nn
&&\hskip30mm\times
\exp\left(2\pi ki\sum_{i=1}^{\ell-1}\sum_{j=1}^{i}\la_j\til{\la}_i
+2\pi ki\sum_{i=1}^\ell\la_i\tsi_\ell\right)~.
\end{eqnarray}
Since the variable $\tsi_\ell$ appears  only in the exponent,
the $\tsi_\ell$ integral is just a $\cob$-function
\begin{equation}
\int d\tsi_\ell 
\exp\left(2\pi ki\sum_{i=1}^\ell\la_i\tsi_\ell\right)
=\frac{1}{ k}\cob\left(\sum_{i=1}^\ell\la_i\right)~.
\end{equation}
After integrating out $\la_\ell$ 
by setting $\la_\ell=-\sum_{i=1}^{\ell-1}\la_i$
by the above $\cob$-function, we get
\begin{eqnarray}
A_{\ell,k}&=&\frac{1}{ k}\int d^{\ell-1}\la d^{\ell-1}\til{\la}
\prod_{i=1}^{\ell-1}\frac{1}{\cosh\pi\la_i\cosh\pi\til{\la}_i}\cdot
\frac{1}{ \cosh\pi(\sum_{i=1}^{\ell-1}\la_i)\cosh\pi(\sum_{i=1}^{\ell-1}\til{\la}_i)}\nn
&&\hskip25mm\times
\exp\left(2\pi ki\sum_{i=1}^{\ell-1}\sum_{j=1}^{i}\la_j\til{\la}_i\right)~.
\label{Ared}
\end{eqnarray}
\subsection{$A_{1,k}$ and $A_{2,k}$}
Let us look closely at the expression (\ref{Ared}) 
for $\ell=1,2$. For $\ell=1$, there is no integral
and the result is simply 
\begin{equation}
A_{1,k}=\frac{1}{ k}~.
\end{equation}
For $\ell=2$, the original four-variable integral
is reduced to a two-variable integral
\begin{equation}
A_{2,k}=\frac{1}{ k}\int d\la d\til{\la}\frac{e^{2\pi ki \la\til{\la}}
}{\cosh^2\pi \la\cosh^2\pi \til{\la}}~.
\end{equation}
The $\til{\la}$-integral can be done by closing the contour 
in the upper half plane when $\la>0$, or the lower half plane when
$\la<0$, and the result turns out to be independent
of the sign of $\la$
\begin{equation}
A_{2,k}=\int d\la\frac{1}{\cosh^2\pi\la}\frac{2\la}{\sinh \pi k\la}~.
\label{A2cosh}
\end{equation} 
Using the relation $\frac{1}{\cosh^2\pi\la}=1-\tanh^2\pi\la$, 
(\ref{A2cosh}) can be further rewritten as
\begin{eqnarray}
A_{2,k}&=&\int d\la\frac{2\la}{\sinh\pi k\la}
-\int d\la\frac{2\la}{\sinh\pi k\la}\tanh^2\pi\la\nn
&=&\frac{1}{ k^2}-\int d\la\frac{2\la}{\sinh\pi k\la}\tanh^2\pi\la~.
\end{eqnarray}

The partition function $Z_{N,k}$ is given by a combination
of $A_{\ell,k}$ (\ref{Zincycle}).
For the $N=2$ case, we find
\begin{equation}
Z_{2,k}=\hf\left[(A_{1,k})^2-A_{2,k}\right]=\int_{-\infty}^\infty
d\la \frac{\la}{\sinh\pi k\la}\tanh^2\pi\la~.
\label{Ztwotanh}
\end{equation}
Note that from (\ref{ladef}) the variable $\la$ corresponds to $\si_1-\tsi_1$.

\section{Evaluation of $Z_{2,k}$}

\begin{figure}[t]
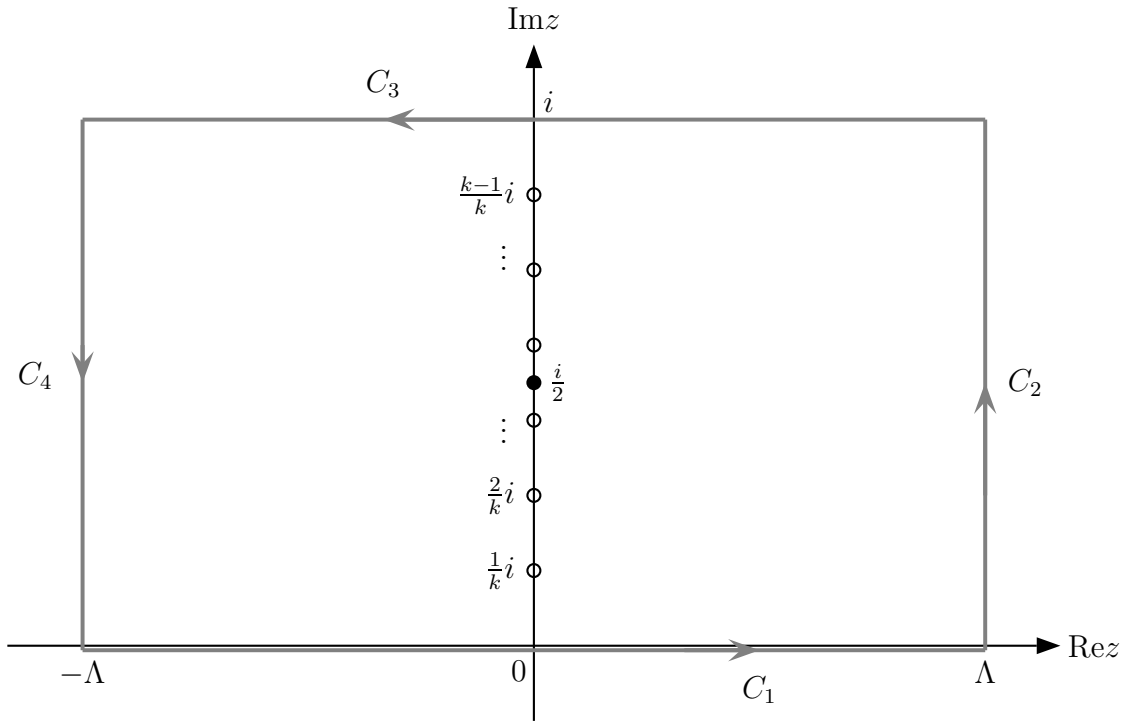

\begin{center}
\pspicture(-8,-1)(8,9)
\psline[linewidth=0.8pt,arrowsize=4pt 3,arrowinset=0]{->}(-7,0)(7,0)
\psline[linewidth=0.8pt,arrowsize=4pt 3,arrowinset=0]{->}(0,-1)(0,8)
\psline[linewidth=1.5pt,linecolor=gray,arrowsize=4pt 3]{->}(-6,-0.06)(3,-0.06)
\psline[linewidth=1.5pt,linecolor=gray](2,-0.06)(6,-0.06)
\psline[linewidth=1.5pt,linecolor=gray,arrowsize=4pt 3]{->}(6,-0.06)(6,3.5)
\psline[linewidth=1.5pt,linecolor=gray](6,2)(6,7)
\psline[linewidth=1.5pt,linecolor=gray,arrowsize=4pt 3]{->}(6,7)(-2,7)
\psline[linewidth=1.5pt,linecolor=gray](0,7)(-6,7)
\psline[linewidth=1.5pt,linecolor=gray,arrowsize=4pt 3]{->}(-6,7)(-6,3.5)
\psline[linewidth=1.5pt,linecolor=gray](-6,4)(-6,-0.06)
\rput[l](7.1,0){Re$z$}
\rput[b](0,8.2){Im$z$}
\rput[t](3,-0.4){$C_1$}
\rput[l](6.3,3.5){$C_2$}
\rput[b](-2,7.3){$C_3$}
\rput[r](-6.4,3.6){$C_4$}
\rput[t](-0.2,-0.2){$0$}
\rput[b](0.2,7.1){$i$}
\pscircle*(0,3.5){0.1}
\rput[l](0.2,3.5){$\frac{i}{2}$}
\pscircle(0,1){0.1}
\pscircle(0,2){0.1}
\pscircle(0,3){0.1}
\pscircle(0,4){0.1}
\pscircle(0,5){0.1}
\pscircle(0,6){0.1}
\rput[r](-0.25,1){$\frac{1}{k}i$}
\rput[r](-0.25,2){$\frac{2}{k}i$}
\rput[r](-0.25,6){$\frac{k-1}{k}i$}
\rput[b](-0.4,5){$\vdots$}
\rput[b](-0.4,2.7){$\vdots$}
\rput[t](6,-0.2){$\Lambda$}
\rput[t](-6,-0.2){$-\Lambda$}
\endpspicture
\end{center}
\caption{This is the contour $C=C_1+C_2+C_3+C_4$ used 
in Appendix B.
$C_1$ and $C_3$ are the horizontal lines at ${\rm Im}z=0$ and ${\rm Im}z=1$, 
respectively.
$C_2$ and $C_4$ are the vertical segments at $|{\rm Re}z|=\Lambda$, and we will 
take the limit $\Lambda\riya\infty$ at the end of computation.}
\label{contour}
\end{figure}

In this Appendix, we evaluate the partition function of 
the $U(2)_k\times U(2)_{-k}$ ABJM theory found 
in Appendix A (\ref{Ztwotanh})
\begin{equation}
Z_{2,k}=\int_{-\infty}^\infty
d\la \frac{\la}{\sinh\pi k\la}\frac{\sinh^2\pi\la}{\cosh^2\pi\la}~.
\label{Zshoch}
\end{equation}
In order to evaluate this integral, we consider 
a contour integral of some holomorphic functions
along the contour $C=C_1+C_2+C_3+C_4$ depicted in Figure 1.
We will specify the relevant holomorphic functions shortly,
which are closely related to the integrand of (\ref{Zshoch}).
It turns out that there are two types of poles inside $C$: the first type
is the poles of $\frac{1}{\sinh\pi kz}$,
and the second type is 
the pole of $\frac{1}{\cosh^2\pi z}$.
We call those poles 
the ``sh-type'' poles and the ``ch-type'' pole, respectively.
Namely,
\begin{eqnarray}
{\rm sh-type~~poles}:&\sinh\pi kz=0~~\riya &z=\frac{s}{ k}i~~(s=1,\cdots,k-1)~,\nn 
{\rm ch-type~~pole}:& \cosh\pi z=0~~\riya &z=\frac{i}{2}~.
\label{listpoles}
\end{eqnarray}
When $k$ is odd, the  pole $z=\frac{i}{2}$ does not appear in the set
of sh-type poles. On the other hand, when $k$ is even,
$z=\frac{i}{ 2}$ is also a pole of $\frac{1}{\sinh\pi kz}$. 
Therefore, we have to analyze the even $k$ case and the odd $k$ case 
separately.

\subsection{Odd $k$ Case}
Let us consider the odd $k$ case first.
In order to evaluate the integral
(\ref{Zshoch}), we introduce the holomorphic function 
\begin{equation}
f(z)=\frac{1}{2}\frac{z-\frac{i}{2}}{\sinh\pi kz}\frac{\sinh^2\pi z}{\cosh^2\pi z}~.
\label{odd-f}
\end{equation}
Since $f(z)$ is regular at $z=0$ and $z=i$, there is no pole on
the contour $C$. Note also that all poles in (\ref{listpoles}) 
are simple poles of $f(z)$. In particular, $z=\frac{i}{2}$ is a simple pole
of $f(z)$ due to the factor of $z-\frac{i}{2}$ in (\ref{odd-f}).

In the contour integral of $f(z)$ along $C$, the contributions
of $C_2$ and $C_4$ become zero in the limit of $\La\riya\infty$
\begin{equation} 
\lim_{\La\riya\infty}\int_{C_2}dzf(z)=
\lim_{\La\riya\infty}\int_{C_4}dzf(z)=0~.
\label{C2C4int}
\end{equation}

As for the integral
along $C_1$ and $C_3$, one can easily see that the limit $\Lambda\riya\infty$ 
exists and leads to a finite result.
Hence, in what follows we will not indicate the limit $\Lambda\riya\infty$
explicitly and we will only write the result of $\Lambda=\infty$.  
The integral along $C_1$ is
\begin{equation}
\int_{C_1}dzf(z)=\int_{-\infty}^\infty d\la f(\la)=
\int_{-\infty}^\infty d\la 
\frac{1}{2}\frac{\la-\frac{i}{2}}{\sinh\pi k\la}
\frac{\sinh^2\pi \la}{\cosh^2\pi \la}=
\int_{-\infty}^\infty d\la 
\frac{1}{2}\frac{\la}{\sinh\pi k\la}\frac{\sinh^2\pi \la}{\cosh^2\pi \la}~.
\end{equation}
Here the term proportional to $-\frac{i}{2}$ vanishes, because the integrand
of that term
is an odd function of $\la$. Therefore, we find
\begin{equation}
\int_{C_1}dzf(z)=\hf Z_{2,k}~.
\label{C1int}
\end{equation}
For the integral along $C_3$, we parametrize $z$ as
\begin{equation}
z=i-\la\quad(\la\in\mathbb{R},-\infty<\la<\infty)~.
\end{equation}
Using the following property of the function $f(z)$
\begin{equation}
f(i-\la)=-f(\la)~,
\end{equation}
we find
\begin{equation}
\int_{C_3}dzf(z)=-\int_{-\infty}^\infty d\la f(i-\la)
=\int_{-\infty}^\infty d\la f(\la)
=\hf Z_{2,k}~.
\label{C3int}
\end{equation}
Combining (\ref{C2C4int}),(\ref{C1int}) and (\ref{C3int}),
we find that 
the partition function $Z_{2,k}$ is equal to the 
integral $\oint_C dz f(z)$
\begin{equation}
\oint_Cdzf(z)=
\sum_{a=1}^4\int_{C_a}dzf(z)
=Z_{2,k}~.
\end{equation}
On the other hand, by the Cauchy's residue theorem
this integral $\oint_C dz f(z)$ can be written as
a sum of residues of the poles inside $C$
\begin{equation}
\oint_Cdzf(z)=2\pi i{\rm Res}_{z=\frac{i}{2}}f(z)+2\pi i\sum_{s=1}^{k-1}
{\rm Res}_{z=\frac{s}{ k}i}f(z)~.
\end{equation}

Putting everything together, we arrive at our final result
\begin{equation}
Z_{2,k}=Z_{2,k}^{({\rm sh})}+Z_{2,k}^{({\rm ch})}~,
\end{equation}
where
\begin{eqnarray}
Z_{2,k}^{({\rm sh})}&=&
2\pi i\sum_{s=1}^{k-1}
{\rm Res}_{z=\frac{s}{ k}i}f(z)=\frac{1}{ k}
\sum_{s=1}^{k-1}(-1)^{s-1}\left(\hf-\frac{s}{ k}\right)\tan^2\frac{\pi s}{ k}~,\nn
Z_{2,k}^{({\rm ch})}&=&
2\pi i{\rm Res}_{z=\frac{i}{2}}f(z)
=\frac{(-1)^{\frac{k-1}{2}}}{\pi}~.
\end{eqnarray}
In the above expression of $Z_{2,k}^{({\rm sh})}$,
using the symmetry under $s\riya k-s$,
 one can show that the sum over the latter half  of $s\in[\frac{k+1}{2}, k-1]$
is the same as the sum over the first half  of $s\in[1,\frac{k-1}{2}]$.
Therefore, the sh-type part can be written as
the twice of the sum over $s\in[1,\frac{k-1}{2}]$
\begin{equation}
Z_{2,k}^{({\rm sh})}=
\frac{2}{ k}
\sum_{s=1}^{\frac{k-1}{2}}(-1)^{s-1}\left(\hf-\frac{s}{ k}\right)
\tan^2\frac{\pi s}{ k}~.
\end{equation}
\subsection{Even $k$ Case}
Next we consider the even $k$ case.
When $k$ is even, $z=\frac{i}{2}$ is a triple zero of the function
$\sinh\pi k z\cosh^2\pi z$ which appears in the denominator of the integral
(\ref{Zshoch}).
Therefore, in order to make $z=\frac{i}{2}$ a simple pole, 
we consider a function $h(z)$ with a factor $(z-\frac{i}{2})^2$
\begin{equation}
h(z)=
\frac{i}{2}\frac{\left(z-\frac{i}{2}\right)^2}{\sinh\pi kz}
\frac{\sinh^2\pi z}{\cosh^2\pi z}~.
\end{equation}
Let us consider the integral of $h(z)$ along the contour $C$
in Figure 1.
As in the previous subsection, we can see that the contributions
from the vertical segments $C_2, C_4$ vanish
\begin{equation}
\lim_{\La\riya\infty}\int_{C_2}dz \,h(z)
=\lim_{\La\riya\infty}\int_{C_4}dz \,h(z)=0~.
\end{equation}

For the integral along $C_1$, from the parity of the integrand
under $\la\riya-\la$, only the term linear in $\la$
survives
\begin{equation}
\int_{C_1}dz \,h(z)=\int_{-\infty}^\infty d\la \,h(\la)
=\int_{-\infty}^\infty d\la\hf\frac{i\la^2+\la-\frac{i}{4}}{\sinh\pi k\la}
\cdot\frac{\sinh^2\pi \la}{\cosh^2\pi \la}=\hf Z_{2,k}~.
\end{equation}
For the integral along $C_3$, using the property
\begin{equation}
h(i-\la)=-h(\la)
\end{equation}
we find
\begin{equation}
\int_{C_3}dz \,h(z)=-\int_{-\infty}^\infty d\la \,h(i-\la)
=\int_{-\infty}^\infty d\la \,h(\la)=\hf Z_{2,k}~.
\end{equation}
Therefore, 
the integral $\oint_Cdz\,h(z)$ is
equal to the partition function $Z_{2,k}$
\begin{equation}
\oint_C dz \,h(z)=
\sum_{a=1}^4\int_{C_a}dz \,h(z)=
Z_{2,k}~.
\end{equation}

By the 
Cauchy's residue theorem, 
$Z_{2,k}$ is written as a sum of residues
of the poles
inside $C$
\begin{equation}
Z_{2,k}=\oint_C dz \,h(z)=2\pi i\sum_{s=1\,(s\not=\frac{k}{2})}^{k-1}
{\rm Res}_{z=\frac{s}{ k}i}h(z)+2\pi i{\rm Res}_{z=\frac{i}{2}}h(z)~,
\end{equation}
where
\begin{eqnarray}
2\pi i\sum_{s=1\,(s\not=\frac{k}{2})}^{k-1}
{\rm Res}_{z=\frac{s}{ k}i}h(z)&=&\frac{1}{ k}
\sum_{s=1\,(s\not=\frac{k}{2})}^{k-1}(-1)^{s-1}
\left(\hf-\frac{s}{ k}\right)^2\tan^2\frac{\pi s}{ k}~,\nn
2\pi i{\rm Res}_{z=\frac{i}{2}}h(z)
&=&\frac{(-1)^{\frac{k}{2}-1}}{ k\pi^2}~.
\end{eqnarray}
The residue of the pole $z=\frac{i}{2}$ can be 
included in the sum of $s$ as the $s=\frac{k}{2}$ term,
with the understanding of taking the limit
\begin{equation}
\lim_{s\riya\frac{k}{2}}\frac{1}{ k}(-1)^{s-1}\left(\hf-\frac{s}{ k}\right)^2
\tan^2\frac{\pi s}{ k}=\frac{(-1)^{\frac{k}{2}-1}}{ k\pi^2}~.
\end{equation} 
Since this term scales as $k^{-1}$,
it seems natural to identify this term
as a part of sh-type contribution.
Therefore, one can think that the partition function for even $k$
consists solely of the sh-type part
\begin{equation}
Z_{2,k}=\frac{1}{ k}
\sum_{s=1}^{k-1}(-1)^{s-1}\left(\hf-\frac{s}{ k}\right)^2\tan^2\frac{\pi s}{ k}~.
\end{equation}
As in the case of odd $k$, the sum over $s$
can be reduced to the half range by using the symmetry
under $s\riya k-s$
\begin{equation}
Z_{2,k}=
\frac{2}{ k}
\sum_{s=1}^{\frac{k}{2}-1}(-1)^{s-1}\left(\hf-\frac{s}{ k}\right)^2
\tan^2\frac{\pi s}{ k}
+\frac{(-1)^{\frac{k}{2}-1}}{ k\pi^2}~.
\end{equation}

\subsection{Some Examples of $Z_{2,k}$ for Low $k$'s}
To see the behavior of the partition function of
$U(2)_k\times U(2)_{-k}$
ABJM theory, here we list
the values of $Z_{2,k}$ from $k=1$ to $k=8$
\begin{align*}
Z_{2,1}&=\frac{1}{\pi} &Z_{2,2}&=\frac{1}{ 2\pi^2}\\
Z_{2,3}&=\frac{1}{3}-\frac{1}{\pi}&Z_{2,4}&=\frac{1}{32}-\frac{1}{4\pi^2} \\
Z_{2,5}&=\frac{10-8\rt{5}}{25}+\frac{1}{\pi} &Z_{2,6}&=-\frac{5}{324}+\frac{1}{6\pi^2} \\
Z_{2,7}&=\frac{5\tan^2\frac{\pi}{7}-3\tan^2\frac{2\pi}{7}+\tan^2\frac{3\pi}{7}}{49} -\frac{1}{\pi}&Z_{2,8}&=\frac{13-8\sqrt{2}}{128}-\frac{1}{8\pi^2}
\end{align*}
It is curious to observe that the
orbifold part of $Z_{2,5}$ and $Z_{2,8}$
are not rational numbers,
and $Z_{2,7}$ cannot be written in a simple form
as a combination of the elementary functions of $k(=7)$,
such as $k^\al$ with some power $\al$.
It would be nice to find a closed form expression of $Z_{2,k}$
as a function of $k$.

%%%%%%%%%%%%%%%%%%%%%%%%%%%%%%%%%%%%%%%%
%\newpage

%%arXiv:cond-mat/9805096 for even/odd N difference


\vskip10mm
\begin{thebibliography}{999}
\parskip=-3pt

%\cite{Aharony:2008ug}
\bibitem{Aharony:2008ug}
  O.~Aharony, O.~Bergman, D.~L.~Jafferis, J.~Maldacena,
  ``N=6 superconformal Chern-Simons-matter theories, M2-branes and their gravity duals,''
  JHEP {\bf 0810}, 091 (2008).
  [arXiv:0806.1218 [hep-th]].

%\cite{Kapustin:2009kz}
\bibitem{Kapustin:2009kz}
  A.~Kapustin, B.~Willett, I.~Yaakov,
  ``Exact Results for Wilson Loops in Superconformal Chern-Simons Theories with Matter,''
  JHEP {\bf 1003}, 089 (2010).
  [arXiv:0909.4559 [hep-th]].

%\cite{Drukker:2010nc}
\bibitem{Drukker:2010nc}
  N.~Drukker, M.~Marino, P.~Putrov,
  ``From weak to strong coupling in ABJM theory,''
  Commun.\ Math.\ Phys.\  {\bf 306}, 511-563 (2011).
  [arXiv:1007.3837 [hep-th]].

%\cite{Drukker:2011zy}
\bibitem{Drukker:2011zy}
  N.~Drukker, M.~Marino, P.~Putrov,
  ``Nonperturbative aspects of ABJM theory,''
  [arXiv:1103.4844 [hep-th]].

%\cite{Fuji:2011km}
\bibitem{Fuji:2011km}
  H.~Fuji, S.~Hirano, S.~Moriyama,
  ``Summing Up All Genus Free Energy of ABJM Matrix Model,''
  JHEP {\bf 1108}, 001 (2011).
  [arXiv:1106.4631 [hep-th]].



%\cite{Santamaria:2010dm}
\bibitem{Santamaria:2010dm}
  R.~C.~Santamaria, M.~Marino, P.~Putrov,
  ``Unquenched flavor and tropical geometry in strongly coupled Chern-Simons-matter theories,''
  [arXiv:1011.6281 [hep-th]].

%\cite{Herzog:2010hf}
\bibitem{Herzog:2010hf}
  C.~P.~Herzog, I.~R.~Klebanov, S.~S.~Pufu and T.~Tesileanu,
  ``Multi-Matrix Models and Tri-Sasaki Einstein Spaces,''
  Phys.\ Rev.\  D {\bf 83}, 046001 (2011)
  [arXiv:1011.5487 [hep-th]].
  %%CITATION = PHRVA,D83,046001;%%

%\cite{Moore:1998et}
\bibitem{Moore:1998et}
  G.~W.~Moore, N.~Nekrasov, S.~Shatashvili,
  ``D particle bound states and generalized instantons,''
  Commun.\ Math.\ Phys.\  {\bf 209}, 77-95 (2000).
  [hep-th/9803265].

%\cite{Kazakov:1998ji}
\bibitem{Kazakov:1998ji}
  V.~A.~Kazakov, I.~K.~Kostov, N.~A.~Nekrasov,
  ``D particles, matrix integrals and KP hierarchy,''
  Nucl.\ Phys.\  {\bf B557}, 413-442 (1999).
  [hep-th/9810035].

%\cite{Pestun:2007rz}
\bibitem{Pestun:2007rz}
  V.~Pestun,
  ``Localization of gauge theory on a four-sphere and supersymmetric Wilson loops,''
   [arXiv:0712.2824 [hep-th]].


%\cite{Kapustin:2010xq}
\bibitem{Kapustin:2010xq}
  A.~Kapustin, B.~Willett, I.~Yaakov,
  ``Nonperturbative Tests of Three-Dimensional Dualities,''
  JHEP {\bf 1010}, 013 (2010).
  [arXiv:1003.5694 [hep-th]].

%\cite{Jafferis:2010un}
\bibitem{Jafferis:2010un}
  D.~L.~Jafferis,
  ``The Exact Superconformal R-Symmetry Extremizes Z,''
  arXiv:1012.3210 [hep-th].
  %%CITATION = ARXIV:1012.3210;%%

%\cite{Hama:2010av}
\bibitem{Hama:2010av}
  N.~Hama, K.~Hosomichi and S.~Lee,
  ``Notes on SUSY Gauge Theories on Three-Sphere,''
  JHEP {\bf 1103}, 127 (2011)
  [arXiv:1012.3512 [hep-th]].
  %%CITATION = JHEPA,1103,127;%%

%\cite{Marino:2011nm}
\bibitem{Marino:2011nm}
  M.~Marino,
  ``Lectures on localization and matrix models in supersymmetric
  Chern-Simons-matter theories,''
  arXiv:1104.0783 [hep-th].
  %%CITATION = ARXIV:1104.0783;%%

%\cite{Suyama:1997ig}
\bibitem{Suyama:1997ig}
  T.~Suyama, A.~Tsuchiya,
  ``Exact results in N(c) = 2 IIB matrix model,''
  Prog.\ Theor.\ Phys.\  {\bf 99}, 321-325 (1998).
  [hep-th/9711073].

%\cite{Green:1996xf}
\bibitem{Green:1996xf}
  M.~B.~Green and M.~Gutperle,
  ``Configurations of two D instantons,''
  Phys.\ Lett.\  B {\bf 398}, 69 (1997)
  [arXiv:hep-th/9612127].
  %%CITATION = PHLTA,B398,69;%%

%\cite{Aharony:2008gk}
\bibitem{Aharony:2008gk}
  O.~Aharony, O.~Bergman and D.~L.~Jafferis,
  ``Fractional M2-branes,''
  JHEP {\bf 0811}, 043 (2008)
  [arXiv:0807.4924 [hep-th]].
  %%CITATION = JHEPA,0811,043;%%

\end{thebibliography}
\end{document}